\documentclass{article}

\usepackage{arxiv}

\usepackage[utf8]{inputenc}      % allow utf-8 input
\usepackage[T1]{fontenc}         % use 8-bit T1 fonts
\usepackage{hyperref}            % hyperlinks
\usepackage{url}                 % simple URL typesetting
\usepackage{booktabs}            % professional-quality tables
\usepackage{amsfonts}            % blackboard math symbols
\usepackage{nicefrac}            % compact symbols for 1/2, etc.
\usepackage{microtype}           % microtypography
\usepackage{graphicx}
\usepackage[numbers]{natbib}
\usepackage{multirow}
\usepackage{titling}
\usepackage{authblk}
\graphicspath{{plots/}} 
\predate{}
\postdate{}

\date{}
\title{From Clicks to Conversations: Evaluating the Effectiveness of Conversational Agents in Statistical Analysis}

\author[1,*]{Qifu Wen}
\author[1]{Prishita Kochhar}
\author[1]{Sherif Zeyada}
\author[2]{Tahereh Javaheri}
\author[1]{Reza Rawassizadeh}

\affil[1]{Department of Computer Science, Boston University Metropolitan College}
\affil[2]{Health Informatics Lab, Boston University Metropolitan College}

\affil[*]{Corresponding author: \texttt{qfwen@bu.edu}}

\begin{document}
\maketitle

\begin{abstract}
The rapid proliferation of data science forced different groups of individuals with different backgrounds to adapt to statistical analysis. We hypothesize that conversational agents are better suited for statistical analysis than traditional graphical user interfaces (GUI). In this work, we propose a novel conversational agent, \emph{\textit{StatZ}}, for statistical analysis. We evaluate the efficacy of \emph{\textit{StatZ}} relative to established statistical software—SPSS, SAS, Stata, and JMP—in terms of accuracy, task completion time, user experience, and user satisfaction. We combined the proposed analysis question from state-of-the-art language models with suggestions from statistical analysis experts and tested with 51 participants from diverse backgrounds. Our experimental design assessed each participant's ability to perform statistical analysis tasks using traditional statistical analysis tools with GUI and our conversational agent. Results indicate that the proposed conversational agents significantly outperform GUI statistical software in all assessed metrics, including quantitative (task completion time, accuracy, and user experience), and qualitative (user satisfaction) metrics. Our findings underscore the potential of using conversational agents to enhance statistical analysis processes, reducing cognitive load and learning curves and thereby proliferating data analysis capabilities, to individuals with limited knowledge of statistics.
\end{abstract}

%% Keywords

\section{Introduction} 
Statistics is one of the main pillars of data science, and data-intensive scientific discoveries have shifted the industry \cite{Hey2009, Kagermann2011}. Statistical analysis stands as a cornerstone in deciphering complex information across various domains, aiding in decision-making and strategic planning. Its pivotal role underscores the necessity for effective tools and methodologies to process and interpret data accurately.

Over the years, numerous software solutions have been developed to facilitate data analysis, especially statistical analysis, with key developments marked by the introduction of foundational tools at various times. SPSS \footnote{\url{https://www.ibm.com/products/spss-statistics}}, one of the earliest statistical analysis tools, launched in 1968, followed by SAS \footnote{\url{https://www.sas.com/en_us/home.html}} in 1976, Stata \footnote{\url{https://www.stata.com/}} in 1985, and JMP \footnote{\url{https://www.jmp.com/en_us/home.html}} in 1989. Apart from programming languages, these tools are still among the most widely used applications for statistical analysis and data engineering \cite{Muenchen2023}.

Despite advancements, the user interfaces of these tools heavily relied on GUI. Although GUI is widely accepted among users, it (i) imposes a high cognitive load on users \cite{Darejeh2024} (ii) requires a longer learning curve than conversational agents \cite{Rawassizadeh2023}, which has led to a decline in user experience. These constraints can hinder their adaptability and accessibility for users with low data science skills and limit their use only for experts. 
   
On the other hand, statistical analysis is required in many fields, from healthcare to economics and Human-Computer Interactions, but research shows that many statistical analyses reported in scientific papers involve mistakes, which have led to flawed results in medicine \cite{Strasak2007}, ecology \cite{Spake2023} and other disciplines.

Since the release of ChatGPT in 2022, Large Language Models (LLMs) have been adopted widely by end users. They can be used for a variety of tasks, including data and statistical analysis. However, they (i) hallucinate and (ii) have limited input token size \cite{brown2020language}. Therefore, they are not well-suited for accurately analyzing tabular data structures, where understanding the relationships among columns is crucial \cite{Zhang2024}. Nevertheless, the simplicity of interacting with conversational agents, the interface used for LLMs, makes them very popular tools that even substitute for web search \cite{xu2023}. 

A conversational interface allows users to navigate and work with ease. This raises the question of how a conversational tool could help reduce the learning curve and increase user satisfaction in data analysis applications. In this research, we propose a novel framework, \emph{\textit{StatZ}}, that focuses on statistical analysis through a conversational agent. During quantitative and qualitative studies, we measure the differences between using a conversational agent and GUIs for statistical analysis. In particular, we use two groups of metrics, qualitative and quantitative, to evaluate task accuracy, task completion time, mouse movement, keyboard clicks, Nielsen's Heuristic, and qualitative user feedback. We found using conversational agents instead of GUI will lead to the wide adoption of statistical analysis tools in the industry. 

\section{Related Work}
Extensive research has explored the effects of conversational agents, particularly their impacts on user experience. Our work underscores a significant evolution in how statistical analyses are performed and how developers interact with software tools. While statistical software has been crucial in data analysis, conversational agents are widely used in customer service, healthcare, and education \cite{Folstad2017}. They aim to enhance user productivity through simplified and natural language interaction. In this section we provide a comparative approach to identify and quantify the specific contributions of conversational agents to user satisfaction and operational efficiency.
\subsection{Conversational Agent Applications}
A wide range of current studies investigates how conversational agents are applied successfully in a variety of contexts.\cite{catania2023conversational} In \textit{customer service}, They have proven scalable for improving content curation and aligning with user needs \citep{candello2022}, including machine-teaching strategies that lower the barriers to conversational agent adoption in areas like banking and telecommunications. 

In \textit{healthcare}, researchers have developed tools such as conversational agents to deliver social and emotional support to patients \citep{wang2021, he2023conversational}, while wearable systems such as CommSense enhance patient-clinician interactions by integrating conversational data analytics \cite{wang2024}. Recent advances in large language models (LLMs) further demonstrate their potential to evaluate and improve communication quality in palliative care and HIV mHealth interactions, offering actionable feedback to enhance rapport and empathy \citep{Wang2025, Wang2024chi}. Meanwhile, \textit{public health interventions} leverage AI-driven messaging to boost the persuasiveness and effectiveness of campaigns such as pro-vaccination efforts \cite{karinshak2023}. 

In the field of \textit{education}, conversational agents help refine course evaluations (EVA) and facilitate informal learning—Design. For instance, Quizzer \cite{schmitt2022} structures community feedback to improve visual design skills\cite{peng2024}, and DebateBot \cite{kim2021} fosters structured, collaborative discussions in classrooms \cite{wambsganss2022}. 

For \textit{group collaboration}, multi-agent platforms such as CommunityBots \cite{Jiang2023} and moderator-focused conversational agents facilitate balanced participation, fairness, and improved decision-making \cite{Do2022} and moderator-focused conversational agents\cite{ bagmar2022}, leading to higher user engagement\cite{kim2021}, better response quality\cite{Do2023}, and fewer conversational disruptions compared to single-agent approaches \citep{Jiang2023}. 

In \textit{design and creativity}, cycles—ProtoChat \cite{choi2021} supports iterative feedback and integrates crowd responses to enhance chatbot scripts, and DesignQuizzer \cite{peng2024} guides novice users in applying visual design principles. Finally, in \textit{advisory services}, multi-party conversational agents augment workflows by contributing social presence and adaptive feedback to bolster user trust and perceived competence, thereby improving both client satisfaction and the advisor's professional standing \citep{bucher2024, schmid2022}.

\subsection{Conversational Agent Impact on User Satisfaction}
Recent studies underscore the multifaceted benefits of conversational agents across several dimensions. With respect to \textit{user satisfaction}, conversational agents that follow user-centered design principles have exhibited significant improvements in user enjoyment and trust \citep{schmitt2022}, higher response quality \citep{wambsganss2022}, and enhanced perceived intelligence through explanation strategies \citep{Do2023}. Furthermore, incorporating conversational repair approaches---strategies for agent error resolution that enable user-initiated corrections, clarifications, or challenges—--can alleviate the impact of false-positive errors and thereby improve the discussion experience \citep{Do2022}, while certain response styles can foster deeper engagement \citep{cho2020}, and even paradoxically, a metaphor signaling lower competence can heighten user satisfaction \citep{khadpe2020}. 

In terms of \textit{efficiency and effectiveness}, conversational agents that pose automatic follow-up questions reduce dropout rates and elicit more informative responses \citep{Hu2024}. They have also demonstrated their potential to gather higher-quality input in course evaluations \cite{wambsganss2022} and group moderation settings \cite{bagmar2022}. Also, multi-agent platforms enhance user engagement and input quality \cite{Jiang2023}, along with efficiency, by reducing context switching and cognitive load \cite{Luger2016}. 

To bolster \textit{user trust and acceptance}, researchers emphasize user-centered designs \citep{schmitt2022, Amershi2019}. For example, learning by teaching paradigms for crowdworkers \citep{Chhibber2022}, and strategies such as algorithmic explanation and effective error-repair \citep{Do2023, Do2022}, while a strong social presence can further elevate perceived competence \citep{schmid2022}. 

Lastly, from a \textit{design and implementation} standpoint, measuring productivity in software projects involves assessing code quality, development speed, and developer satisfaction \citep{McConnell2006, Sadowski2015}, prompting researchers to explore alternative metrics. Within these workflows, refining question-asking techniques \citep{Hu2024} and integrating crowd feedback into conversation design \citep{choi2021} have proven fruitful, as has employing AI-generated text to improve message persuasiveness \citep{karinshak2023}. Studies have shown that conversational tools can also enhance the developer experience by providing immediate, context-aware support \citep{Wang2018}. 

Collectively, these findings reveal that well-designed, context-aware conversational agent can substantively enhance user satisfaction, efficiency, trust, collaboration, and design outcomes. Despite these promising efforts, we didn't find a work that assists in statistical analysis and data engineering with a conversational agent, which is the focus of this research.

\section{Methods}
We designed a novel framework for statistical analysis, and we assessed the usability and efficiency of our approach against traditional statistical analysis tools by recruiting participants with varying levels of statistical familiarity. This study aims to provide insights into each tool's usability, guiding potential software enhancements and improving user experience in statistical data analysis.

Our study is IRB-approved, and to ensure users' privacy and confidentiality, all users' identification data were deleted at the conclusion of the study.

\subsection{Tasks}
To systematically identify the most common tasks performed by statisticians and data scientists and incorporate them into our framework, we utilized three different language models: ChatGPT-4o\footnote{OpenAI. “GPT-4.” 2023. \url{https://openai.com/product/gpt-4}. Accessed June 2024.}, 
Meta's Llama 405B model\footnote{Touvron, Hugo, et al. “LLaMA: Open and Efficient Foundation Language Models.” 
\textit{arXiv preprint} arXiv:2302.13971 (2023). \url{https://arxiv.org/abs/2302.13971}. Accessed June 2024.}, 
and Claude Sonte v3.5\footnote{Anthropic. “Claude.” 2023. \url{https://www.anthropic.com/index/introducing-claude}. Accessed June 2024.}. These models were prompted to generate lists of 10 common statistical tasks on tabular data. Subsequently, an expert user, who is a professor of statistics, shortlists common tasks that were most frequently repeated across all model responses and are not related to machine learning. For example, some LLMs consider logistic regression to be statistical analysis, and this has been removed from the list of tasks. 

Participants were asked to perform a series of ten tasks using each of the five software tools. A detailed description of tasks are provided in the appendix. These tasks were designed to assess their ability to utilize fundamental and commonly used statistical functions within each software, focusing on data manipulation, statistical computations, and visualization techniques. 
For data analysis, participants were provided with two datasets: the Iris Dataset  \url{https://archive.ics.uci.edu/ml/datasets/Iris} and the NYC Taxi Dataset \footnote{ \url{https://archive.ics.uci.edu/ml/datasets.html}}. Both datasets, are baseline datasets for machine learning and statistical analysis and both datasets reflect real-world analytical demands.
While the specific tasks varied slightly between the Iris and NYC Taxi datasets to suit their respective data characteristics, the overall structure and objectives of the tasks remained consistent across datasets and software tools to allow for comparative analysis of usability and efficiency.

 \subsection{Implementation and Design}
\begin{figure}[h!]
    \centering
    \includegraphics[width=1\textwidth]{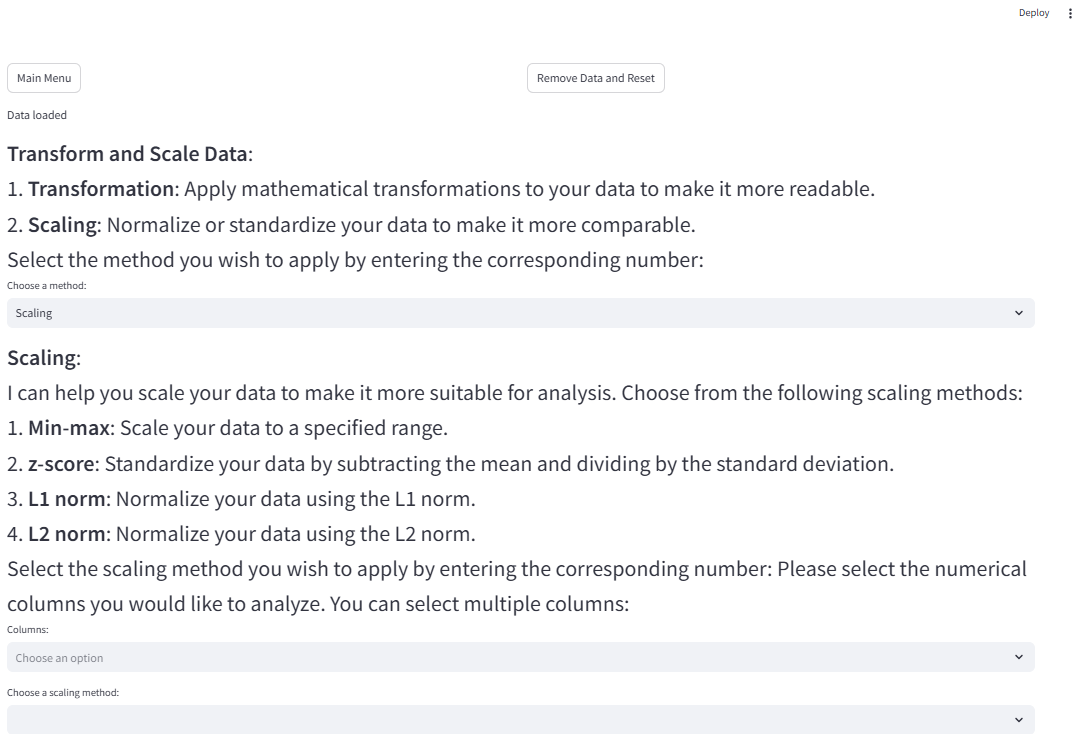}
    \caption{Front-end sample for transform and scaling data. User can select the method of their choice from Min-max scaling, z-score scaling, L1 norm scaling and L2 norm scaling.}
    \label{fig:frontend_scaling}
\end{figure}

To implement the conversational agent with statistical capabilities, we have used Python v3.9\footnote{Python Software Foundation. \textit{Python Language Reference, version 3.9}. Retrieved from \url{https://www.python.org}}  for the back end, and Streamlit v 1.4 \footnote{Streamlit. \textit{Streamlit: The fastest way to build data apps}. Retrieved from \url{https://streamlit.io}} for the front end. Statistical libraries used in the backend include Scikit-learn \cite{pedregosa2011scikit}, SciPy\cite{2020SciPy-NMeth}, Pandas\cite{mckinney2010data}, NumPy\cite{harris2020array} and Plotly\footnote{Plotly Technologies Inc. 
    \textit{Collaborative data science}. 
    Retrieved from \url{https://plotly.com}}.
Figure \ref{fig:frontend_scaling} depicts the framework's interactive interface, featuring two sample interfaces for scaling selection, augmented with contextual guidance and systematic decision pathways to ensure informed user choices.

\subsection{Participants}

The study recruited participants from diverse backgrounds to assess the usability of statistical analysis tools among users with varying levels of statistical proficiency. Participants self-reported their familiarity with statistical concepts, such as hypothesis testing, probability distributions, and data analysis, on a Likert scale from 1 (not familiar) to 5 (very familiar). This measure allowed for the categorization of participants based on their statistical knowledge, ensuring a heterogeneous sample. 

The inclusion criteria were minimal to encourage broad participation. Participants needed access to a laptop and basic proficiency with Windows operating systems, as the virtual machine was Windows-based. Problem-solving skills were required, but prior knowledge of statistical analysis or programming was not necessary.

The study included 51 participants—30 men and 21 women—with no individuals identifying as non-binary, with an age range between 21 to 47 (mean=24, SD=3.5). They were hired through the announcement in four courses of the authors' university. Participants self-assessed their familiarity with statistical concepts across five levels. Level 1 (minimal prior knowledge) had 13 participants (8 men, 5 women). Level 2 included 5 participants (3 men, 2 women). Level 3, representing moderate familiarity, was the largest group, with 14 participants evenly split between men and women. Level 4 had 12 participants (10 men, 2 women), indicating higher proficiency. Level 5, the most proficient group, comprised 7 participants (2 men, 5 women). This diverse distribution reflects a broad spectrum of statistical experience within the study group.

Notably, the number of participants who were "Not Familiar At All" with statistics (13 participants) was almost as high as the most proficient group, highlighting our sample's inclusiveness of complete beginners.  Conversely, 7 participants considered themselves "Very Familiar" with statistics, indicating advanced understanding capable of complex analyses. Their feedback is particularly valuable for assessing the depth and advanced capabilities of the statistical tools studied.

Participants received verbal detailed information about the study for half an hour, including a video guide to set up the necessary software and an explanation of the study's purpose. This guide covered the use of the Remote Desktop—a virtual environment where all testing was conducted—and instructions on running a script to track their activity while respecting their privacy and running it only during the experiment. 

\subsection{Apparatus}

The study was conducted using a remote computer at the authors' university, ensuring all participants operated with the same hardware and software. The experiment machines include on an Intel(R) Xeon(R) Platinum 8370C CPU at 2.80 GHz, with 16.0 GB of RAM. They operate on Windows 11 Enterprise multi-session with a 64-bit system, up-to-date with the latest security and software features as of June 26, 2024. This uniform setup across all users helped maintain consistency over the study's technical environment.

This environment operated on a Windows virtual machine equipped with a range of statistical analysis software, including JMP Pro v. 17, StataSE v.18, SAS v.9.4, IBM SPSS Statistics v. 29.0.1. These Proprietary tools have long been standard in industries requiring rigorous statistical analysis \cite{Raykov2006}. Our conversational agent is deployed on a web page and accessible to participants from the virtual machine. 

Alongside the software setup, a script was utilized to monitor user interactions within the remote desktop environment, tracking the duration of activity, number of keystrokes, mouse clicks, and cursor movement. This tracking was operated only during the study and is used to evaluate participant engagement and interaction with the listed tools. To respect participants privacy after analysis of the data, the track file were discarded.

\subsection{Procedure}

In preparation for the study, participants registered via an online form and configured access to the remote computers using their university credentials. They were asked to ensure that their laptops were fully charged and capable of running remote desktop software.
We employed a balanced Latin square \cite{Mandal2017} for the experiment in order to minimize the learning effect and reduce the carry-over effect among participants \cite{Brooks2012}.

To systematically manage the experimental setup, the study was structured with five participants per session to maintain manageable group sizes and facilitate observation. Each study session is administered by two individuals. Each participant was assigned a specific sequence for utilizing the five statistical software tools to counterbalance potential order effects, as presented in Table~\ref{tab:software_sequence}.

\begin{table}[h]
    \centering
    \caption{Assigned Sequence of Statistical Software Tools per Participant}
    \label{tab:software_sequence}
    \begin{tabular}{|c|c|c|c|c|c|}
    \hline
    \textbf{Participant} & \textbf{1st Tool} & \textbf{2nd Tool} & \textbf{3rd Tool} & \textbf{4th Tool} & \textbf{5th Tool} \\ \hline
    1 & JMP Pro 17 & StataSE 18 & SAS 9.4 & IBM SPSS Statistics & \textit{StatZ} \\ \hline
    2 & StataSE 18 & JMP Pro 17 & \textit{StatZ} & IBM SPSS Statistics & SAS 9.4 \\ \hline
    3 & SAS 9.4 & IBM SPSS Statistics & \textit{StatZ} & JMP Pro 17 & StataSE 18 \\ \hline
    4 & IBM SPSS Statistics & SAS 9.4 & JMP Pro 17 & StataSE 18 & \textit{StatZ} \\ \hline
    5 & \textit{StatZ} & JMP Pro 17 & StataSE 18 & SAS 9.4 & IBM SPSS Statistics \\ \hline
    \end{tabular}
\end{table}

For each statistical tool, participants initiated a 20-minute timer to standardize task duration and launched a Python-based activity tracking program within the remote desktop environment to monitor their interactions. 

Participants were permitted to utilize external resources such as Internet searches and AI assistants to aid in completing the tasks. Technical assistance was available for issues related to setting up the remote desktop and the activity tracking program; however, no assistance was provided regarding the use of the statistical software or the execution of the analyses to preserve the integrity of the results.

All participants were instructed not to exceed the 20-minute time limit per software tool to ensure consistency across all sessions. If they completed the tasks before the allotted time, they proceeded to the next software tool after following the prescribed data submission procedures.

\subsection{Objectives}
Our studies aimed to assess the usability of our proposed conversational agent against GUI-based statistical tools, among users with varying levels of expertise. We focus on qualitative and quantitative comparisons across different software environments, yielding insights into the strengths and limitations of graphical user interfaces versus a conversational agent for statistical analysis tasks.

\section{Experiments}
In this section, we list and describe both quantitative and qualitative experiments we conduct to evaluate our approach among traditional statistical tools.

\subsection{Task Completion Accuracy}
To measure the users' accuracy in performing given tasks, we measure the correctness of the result. For example, to compare two unrelated columns of a dataset that do not have a Gaussian distribution, the user should select a non-parametric significance test. We hypothesize that existing GUI tools do not disallow users from selecting the wrong significance test, and a conversational agent can guide the user to the correct significance test. 
After users performed the described tasks in all tools, we graded each incorrect outcome as 0 and the corrected outcome as 1 per task.

\subsection{Mouse and Keyboard Interaction}
A known factor that quantifies user efficiency \cite{Souza2021} is monitoring user activity across three dimensions: keyboard inputs, mouse clicks, and mouse movement distance. This monitoring encompassed all activities during the experiment, including online search for a solution and the use of additional software applications. We hypothesize that these data accurately reflect users' efforts in real-world scenarios.

The efficiency metrics collected from our study enable a comparative analysis of five statistical software packages. These metrics include average task completion time (measured in seconds), average number of keyboard inputs, average number of mouse clicks, and average mouse movement distance (converted from pixels to meters based on a screen resolution of $1024\times768$).

\subsection{Task Completion Time}
Another important factor that can testify to the quality of a software is its task completion time, which has a direct correlation with the cognitive load \cite{longo2018experienced}. We hypothesize that conversational agents reduce the cognitive load because users don't need to explore different options, and this might improve task completion time. Therefore, as another quantitative study, we measure the task completion time.

\subsection{Nielsen's Heuristic Analysis}
To conduct the user experiment study, we evaluated five different software packages across ten usability heuristics adapted from Jakob Nielsen's principles for effective interface design \cite{Nielsen1994}. Participants were asked to rate each software on a scale of 1 to 5 for criteria such as the clarity of feedback, the use of familiar language, error prevention, and the ease of reversing actions. The results were further analyzed to assess how well each software aligned with Nielsen's heuristics, particularly in areas such as user control, consistency, and recognition rather than recall.

\subsection{Qualitative User Feedback}
To gain qualitative insights into user experiences and perceptions of the program, we included an optional open-ended question asking participants whether they enjoyed using the software and what improvements they would like to see. Proposing open-ended questions for qualitative evaluation is inspired by Patton \cite{patton1990qualitative}. With a response rate of 60.7\%, we collected valuable qualitative data for analysis. The responses were analyzed thematically to identify recurring patterns, such as common usability issues, features users found most valuable, and suggestions for enhancing the overall user experience. This approach allowed us to complement the quantitative data with rich, contextual insights, providing a more comprehensive understanding of user satisfaction and areas for improvement.

\subsection{Results}
\subsubsection{Task Completion Accuracy}
To assess the accuracy of responses, their responses were compared against established correct answers. For the given list of tasks among 51 participants, the result of this experiment reveals that \textit{StatZ} achieved the highest average accuracy of 0.8009. Other tools (JMP, SAS, SPSS, and Stata) showed lower accuracies of 0.3100, 0.2852, 0.2962, and 0.4556, respectively. 

\begin{figure}[htb]
  \centering
  \begin{minipage}{0.48\textwidth}
    \centering
    \includegraphics[width=\textwidth]{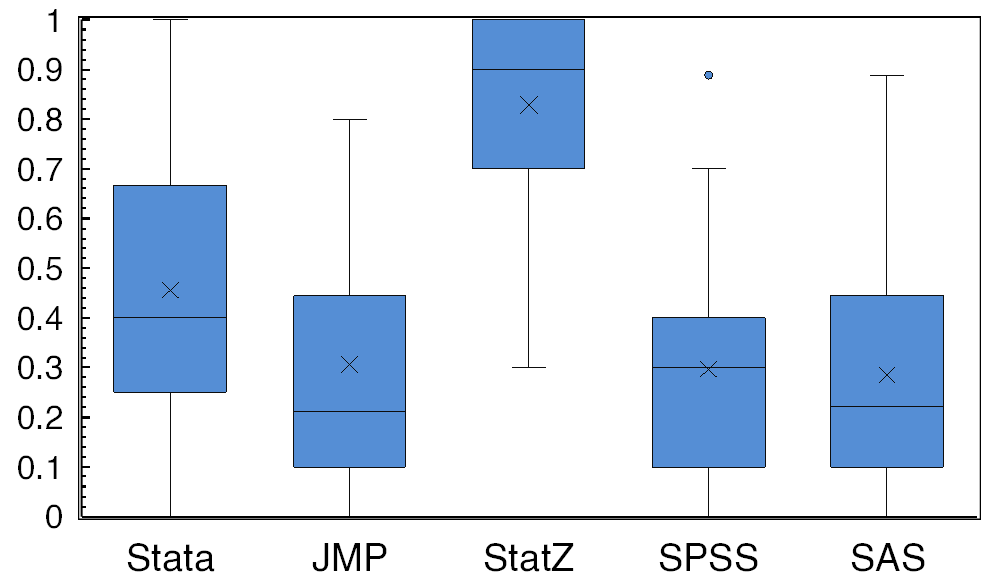}
  \caption{Box Plot of Accuracy Across Different Software}
  \label{fig:accuracy_box_plot}
  \end{minipage}
  \hfill
  \begin{minipage}{0.48\textwidth}
    \centering
    \includegraphics[width=\textwidth]{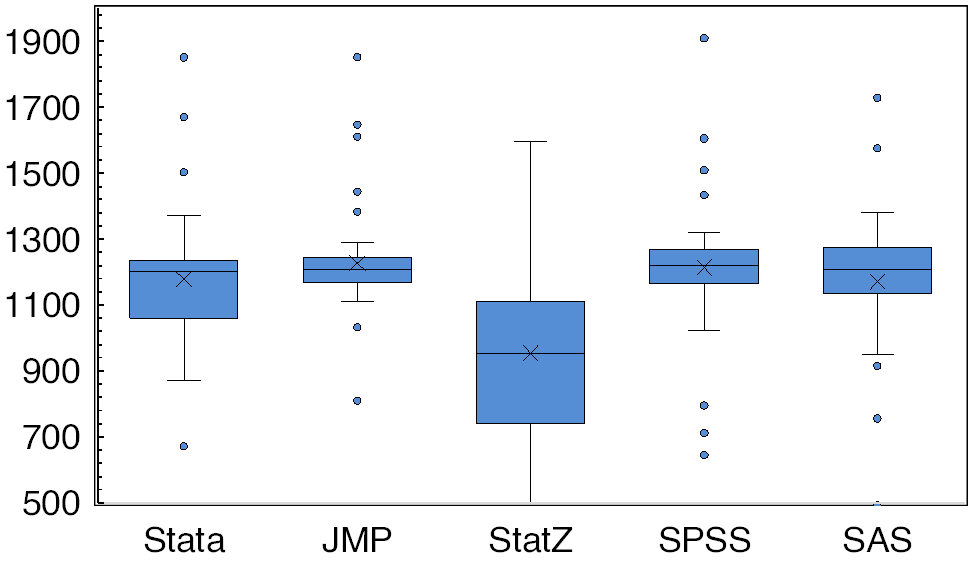}
  \caption{Box Plot of Time Efficiency Using Various Statistical Software}
  \label{fig:time_box_plot}
  \end{minipage}
\end{figure}

To ensure we observe statistically significant differences between \textit{StatZ} and other groups, we employed a statistical significance test. Initially, the Shapiro-Wilk test \cite{Shapiro1965}  was used to assess the normality of accuracy data across different types of statistical software, revealing non-normal distributions for each. Therefore, we adopt a non-parametric significant test. Since we are doing repeated measures (different software for the same subjects), we use the Friedman test \cite{Friedman1937}. The Friedman test output is 83.103 with 4 degrees of freedom, and a high significant p-value (\( p < 3.9 \times 10^{-17} \)), confirming significant disparities in performance among the tested software. Furthermore, the effect size for Friedman Test Kendall's $W = 0.3919934$. 

After we identified at least one group that's different from the rest using the Friedman test, we used the postHoc test to compare the pairwise difference, which is the Nemenyi test \cite{nemenyi1963}. The pairwise Nemenyi comparisons presented in Table \ref{tab:combined_results} (with Bonferroni corrections \cite{Bonferroni1936}) demonstrate statistically significant differences in accuracy across the tested software tools. With all pairwise comparisons involving \textit{StatZ} displaying highly significant \(p\)-values. These results underscore \textit{StatZ}'s superior performance, suggesting that the employment of conversational agents facilitates more accurate statistical analyses. 

In particular, these findings suggest that using a conversational agent leads to higher accuracy compared to the GUI tools. The observed improvement in \textit{StatZ}'s accuracy not only provides a robust indication of conversational agent efficacy but also supports its potential to enhance reliability and user outcomes in practical data analysis scenarios.

\subsubsection{ Mouse and Keyboard Interaction}
\begin{figure}[htb]
  \centering
  \begin{minipage}{0.48\textwidth}
    \centering
    \includegraphics[width=\textwidth]{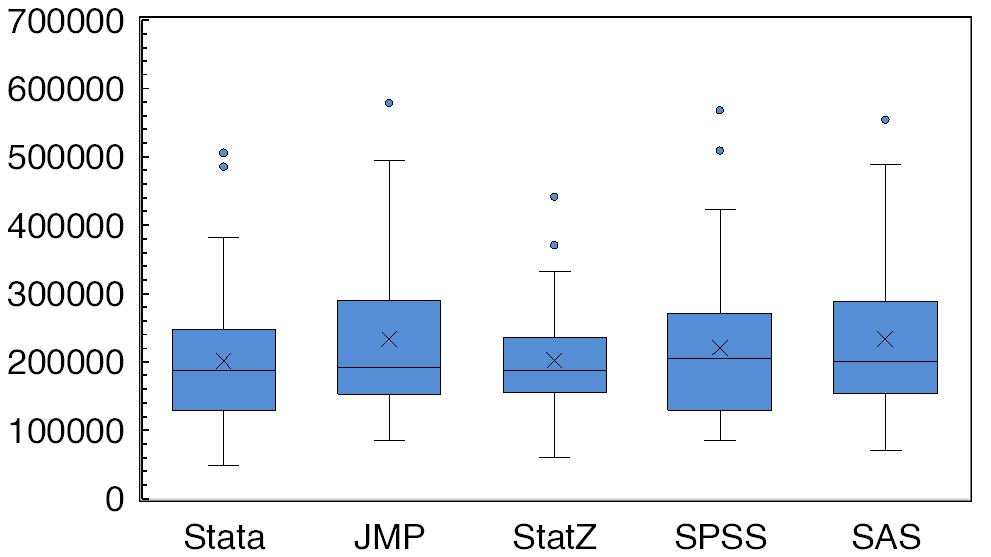}
    \caption{Average Mouse Distance Traveled Among Statistical Tools}
    \label{fig:mouse_distance_plot}
  \end{minipage}
  \hfill
  \begin{minipage}{0.48\textwidth}
    \centering
    \includegraphics[width=\textwidth]{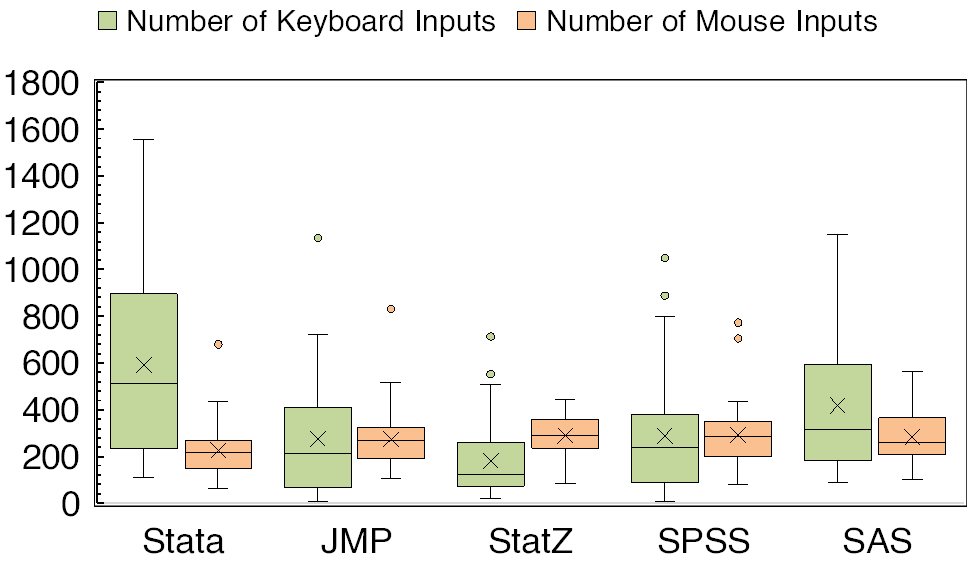}
    \caption{Comparison of Keyboard and Mouse Usage Across Tools}
    \label{fig:keyboard_mouse_plot}
  \end{minipage}
\end{figure}

\textit{StatZ} demonstrated a significantly lower average number of keyboard inputs at 182, compared to JMP (275), SAS (417), SPSS (289), and Stata (593). This substantial reduction indicates that \textit{StatZ} requires fewer keystrokes to perform statistical operations, suggesting a more streamlined or intuitive input process. Additionally, \textit{StatZ} recorded the second-lowest mouse movement distance at 187,283 pixels, slightly above Stata's 186,915 pixels but less than JMP (192,322 pixels), SAS (200,389 pixels), and SPSS (205,120 pixels). Despite having a comparable number of mouse clicks to the other software packages, the reduced mouse movement distance implies that \textit{StatZ}'s user interface facilitates more efficient navigation, potentially through conversational agent layout that minimizes the cursor movement between tasks.

\begin{table}[htbp]
\centering
\caption{Combined Statistical Analysis Results for Accuracy, Time, \# of Keyboard Inputs, \# of Mouse Inputs, Total Distance}
\resizebox{0.9\textwidth}{!}{%
\begin{tabular}{lllrcl}
\toprule
\textbf{Measure} & \textbf{Test} & \textbf{Comparison} & \textbf{Statistic} & \textbf{p-value} & \textbf{Reject H$_{0}$}\\
\midrule

%%======================================================
%% ACCURACY
%%======================================================
\multirow{11}{*}{\textbf{Accuracy}}
 & \multirow{1}{*}{Friedman}
    & All Software & $\chi^2 = 83.1026$ & 3.832e--17 & \textbf{Yes} \\
 \cmidrule(lr){2-6}
 & \multirow{10}{*}{Nemenyi}
    & (JMP vs.\ SAS)    & -- & 8.552e--01 & False \\
 &  & (JMP vs.\ SPSS)   & -- & 9.613e--01 & False \\
 &  & (JMP vs.\ \textit{StatZ})  & -- & 2.041e--09 & \textbf{True} \\
 &  & (JMP vs.\ Stata)  & -- & 4.790e--02 & \textbf{True} \\
 &  & (SAS vs.\ SPSS)   & -- & 9.977e--01 & False \\
 &  & (SAS vs.\ \textit{StatZ})  & -- & 1.858e--12 & \textbf{True} \\
 &  & (SAS vs.\ Stata)  & -- & 1.698e--03 & \textbf{True} \\
 &  & (SPSS vs.\ \textit{StatZ}) & -- & 1.979e--11 & \textbf{True} \\
 &  & (SPSS vs.\ Stata) & -- & 5.633e--03 & \textbf{True} \\
 &  & (\textit{StatZ} vs.\ Stata)& -- & 2.780e--03 & \textbf{True} \\
\midrule

%%======================================================
%% TIME
%%======================================================
\multirow{11}{*}{\textbf{Time}}
 & \multirow{1}{*}{Friedman}
    & All Software & $\chi^2 = 27.0240$ & 1.9658e--05 & \textbf{Yes} \\
 \cmidrule(lr){2-6}
 & \multirow{10}{*}{Nemenyi}
    & (JMP vs.\ SAS)   & -- & 0.902229          & False \\
 &  & (JMP vs.\ SPSS)  & -- & 0.980783          & False \\
 &  & (JMP vs.\ \textit{StatZ}) & -- & 0.000321          & \textbf{True} \\
 &  & (JMP vs.\ Stata) & -- & 0.966180          & False \\
 &  & (SAS vs.\ SPSS)  & -- & 0.598163          & False \\
 &  & (SAS vs.\ \textit{StatZ}) & -- & 0.009699          & \textbf{True} \\
 &  & (SAS vs.\ Stata) & -- & 0.999348          & False \\
 &  & (SPSS vs.\ \textit{StatZ})& -- & 0.000025          & \textbf{True} \\
 &  & (SPSS vs.\ Stata)& -- & 0.744284          & False \\
 &  & (\textit{StatZ} vs.\ Stata)& --& 0.004294          & \textbf{True} \\
\midrule

%%======================================================
%% KEYBOARD INPUTS
%%======================================================
\multirow{11}{*}{\parbox{3.7cm}{\textbf{\# of Keyboard Inputs}}}
 & \multirow{1}{*}{Friedman}
    & All Software & $\chi^2 = 49.4728$ & 4.6524e--10 & \textbf{Yes} \\
 \cmidrule(lr){2-6}
 & \multirow{10}{*}{Nemenyi}
    & (JMP vs.\ SAS)    & -- & 7.056651e--04  & \textbf{True} \\
 &  & (JMP vs.\ SPSS)   & -- & 6.878011e--01  & False \\
 &  & (JMP vs.\ \textit{StatZ})  & -- & 9.999989e--01  & False \\
 &  & (JMP vs.\ Stata)  & -- & 3.747513e--07  & \textbf{True} \\
 &  & (SAS vs.\ SPSS)   & -- & 6.044237e--02  & False \\
 &  & (SAS vs.\ \textit{StatZ})  & -- & 5.816757e--04  & \textbf{True} \\
 &  & (SAS vs.\ Stata)  & -- & 5.371886e--01  & False \\
 &  & (SPSS vs.\ \textit{StatZ}) & -- & 6.584014e--01  & False \\
 &  & (SPSS vs.\ Stata) & -- & 2.625518e--04  & \textbf{True} \\
 &  & (\textit{StatZ} vs.\ Stata)& -- & 2.874311e--07  & \textbf{True} \\
\midrule

%%======================================================
%% MOUSE INPUTS
%%======================================================
\multirow{11}{*}{\parbox{3.7cm}{\textbf{\# of Mouse Clicks}}}
 & \multirow{1}{*}{Friedman}
    & All Software & $\chi^2 = 15.6507$ & 3.5256e--03 & \textbf{Yes} \\
 \cmidrule(lr){2-6}
 & \multirow{10}{*}{Nemenyi}
    & (JMP vs.\ SAS)    & -- & 0.980783       & False \\
 &  & (JMP vs.\ SPSS)   & -- & 0.999730       & False \\
 &  & (JMP vs.\ \textit{StatZ})  & -- & 0.980783       & False \\
 &  & (JMP vs.\ Stata)  & -- & 0.009699       & \textbf{True} \\
 &  & (SAS vs.\ SPSS)   & -- & 0.945592       & False \\
 &  & (SAS vs.\ \textit{StatZ})  & -- & 1.000000       & False \\
 &  & (SAS vs.\ Stata)  & -- & 0.053317       & False \\
 &  & (SPSS vs.\ \textit{StatZ}) & -- & 0.945592       & False \\
 &  & (SPSS vs.\ Stata) & -- & 0.005079       & \textbf{True} \\
 &  & (\textit{StatZ} vs.\ Stata)& -- & 0.053317       & False \\
\midrule

%%======================================================
%% TOTAL DISTANCE
%%======================================================
\multirow{1}{*}{\textbf{Total Distance}}
 & \multirow{1}{*}{Friedman}
    & All Software & $\chi^2 = 5.2549$ & 2.6213e--01 & No \\
\bottomrule
\end{tabular}
} % end of resizebox
\label{tab:combined_results}
\end{table}

\subsubsection{Task Completion Time}
In terms of task completion time, \textit{StatZ} outperforms the other tools significantly. Users spent an average of 954 seconds (approximately 15.9 minutes) to complete 10 tasks, which is considerably lower than the times recorded for JMP (1,270 seconds, approximately 21.2 minutes), SAS (1,172 seconds, approximately 19.5 minutes), SPSS (1,251 seconds, approximately 20.9 minutes), and Stata (1,226 seconds, approximately 20.4 minutes).

Building on these observations, the results from the Friedman test (\(\chi^2 = 27.024\), \(p = 1.965 \times 10^{-5}\)) and subsequent pairwise Nemenyi comparisons (adjusted via Bonferroni correction) confirmed statistically significant differences in completion time across the five software tools. Crucially, \textit{StatZ} demonstrated a distinctly lower average completion time than JMP, SAS, SPSS, and Stata, with pairwise comparisons showing significant \(p\)-values in each case. 

\subsubsection{Nielsen's Heuristic Analysis}

The aggregated of Nielesen's questions indicate that \textit{StatZ} software consistently outperformed other packages in every evaluated category. For instance, \textit{StatZ} scored highest in providing clear and effective feedback with a total of 215 points, compared to its closest competitor, SPSS, which scored 160 points in the same category. Similar trends were observed in other areas such as error prevention and user control, where \textit{StatZ} garnered 191 and 201 points respectively, significantly ahead of other software.

\begin{figure}[htbp]
  \centering
  \includegraphics[width=0.8\textwidth]{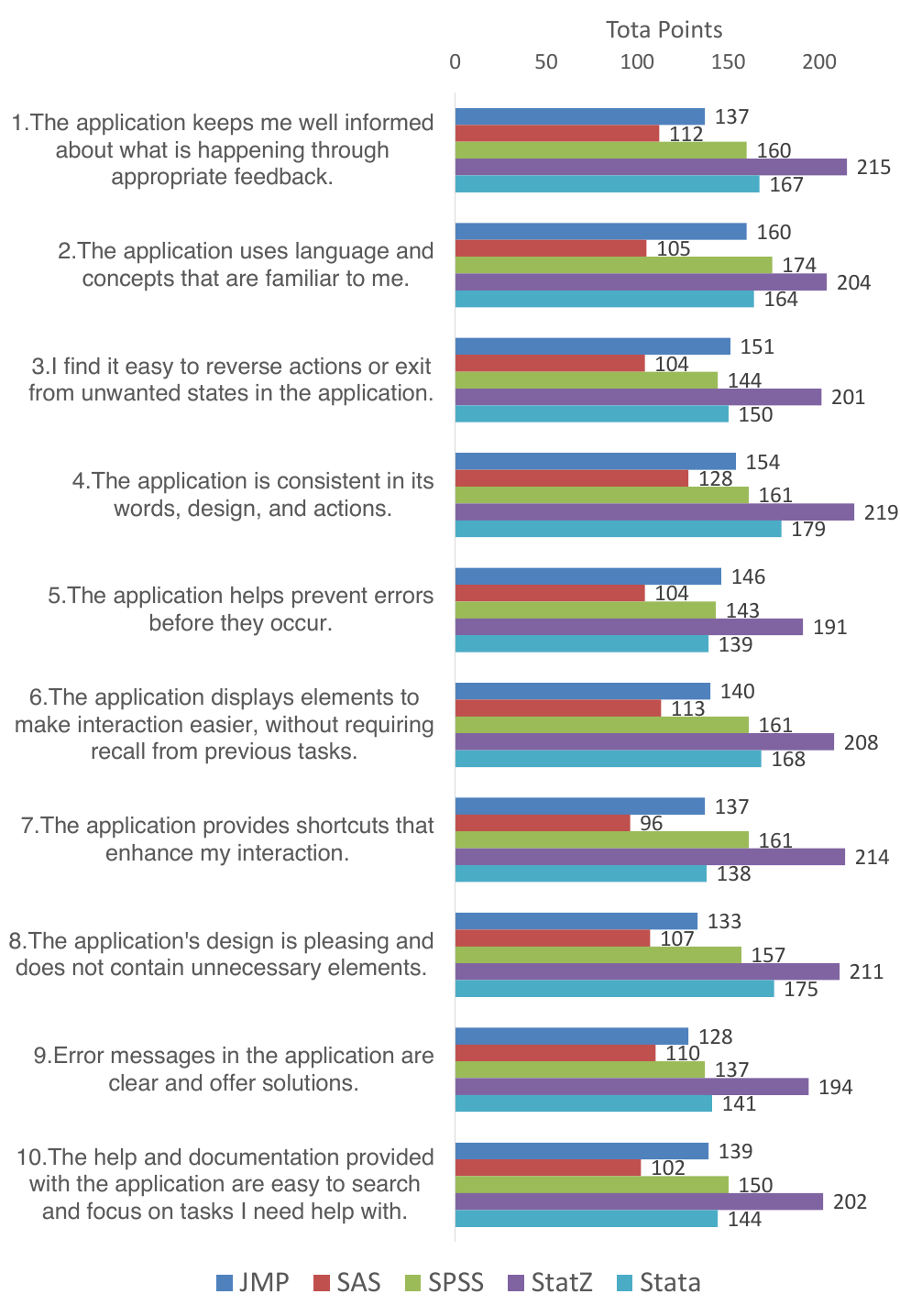}
  \caption{Heuristic Evaluation Summary of Software Usability}
 \label{fig:userfeedback}
\end{figure}

Figure ~\ref{fig:userfeedback} shows the result of this study. 
Participants rated \textit{StatZ} highest in key areas such as clarity of feedback, intuitive language, error prevention, and ease of reversing actions (Figure~\ref{fig:userfeedback}). This detailed plot illustrates \textit{StatZ}'s superiority in facilitating a user-friendly and intuitive interface. In particular,  results present \textit{StatZ} offers a more user-friendly and satisfying experience. The notable differences in user satisfaction between \textit{StatZ} and other tools emphasize the importance of user-centered design in statistical software, and \textit{StatZ}'s success highlights conversational agents potential to increase user engagement and productivity.

\subsubsection{Qualitative User Feedback}
Our qualitative post-study evaluation implies that (i) the guidance provided by the conversational agent and (ii) limiting the user choice leads to higher accuracy and faster response time in performing the given statistical tasks.

Three independent reviewers coded the feedback, and inter-rater reliability was assessed using Fleiss' Kappa. The resulting value of 0.68 indicates good agreement among the raters, enhancing the credibility of our qualitative analysis.

\begin{table}[h!]
\centering
\begin{tabular}{|l|p{10cm}|}
\hline
\textbf{Category} & \textbf{Labels} \\ \hline
\textbf{Learning and Documentation} & Documentation, Learning Curve, Tutorials, Guidance, Help, Onboarding, Quick Tips, Complexity, First-Time User, Detailed Instructions \\ \hline
\textbf{GUI and Usability Improvements} & User-Friendly, Menu, Visualizations, Modernization, Interface Clarity, Visual Navigation, Dropdown Menus, Icons, Interactive UI, Modern GUI, UI Simplicity, User Experience, Intuitive Design, Simplicity, Ease of Navigation \\ \hline
\textbf{User Sentiments and Preferences} 
& Satisfactory, Beneficial, Appreciated Features, Exceptional Experience,  Unhelpful Features, Non-Intuitive, Not User-Friendly, Complexity, Outdated \\ \hline
\textbf{Data Handling and Operations} & Data Import, Data Export, Data Management, Data Visibility, Dataset Visibility, Data Files, Data Upload \\ \hline
\textbf{Performance and Efficiency} & Performance, Efficiency, Response Time, Productivity, Quick Task Completion, Slow Processing, Lagging, Time-Consuming, Slow Response \\ \hline
\textbf{Navigation and Interaction} & Navigation, Interface Navigation, Menu Navigation, User Interaction, Visual Navigation \\ \hline
\textbf{Search and Discovery Features} & Search Function, Discovery Features, Finding Tools, Search Capabilities \\ \hline
\textbf{Error Handling and Feedback} & Errors, Error Handling, Troubleshooting, Error Notifications, Vague Errors, Unhelpful Error Messages, Program Freezes, Crashes,  Stability Issues \\ \hline
\textbf{Customization and Flexibility} & Tailored Functions, Customize, Scaling, Autosave, Command History, Advanced Functionalities, Additional Features \\ \hline
\end{tabular}
\caption{Revised Categories and Labels for Qualitative User Feedback}
\label{tab:revised_categories_labels}
\end{table}

16 participants reported positive feedback with \textit{StatZ}, three participants reported positively with SPSS and JMP each, five participants reported positively with Stata. None of the participants reported positive experience with SAS. 

Based on the negative feedback participants provided, the most frequently mentioned theme across all software was GUI and usability improvements, with a total of 49 mentions. This was followed by Learning and Documentation (31 mentions), highlighting a substantial need for better user interfaces and instructional resources. Positive feedback and preferences were noted 27 times, suggesting that while users had critiques, many also appreciated certain aspects of the software. Negative feedback and preferences appeared 22 times, reflecting specific dissatisfactions.

In the following, we provide qualitative feedbacks analysis for each software in more details:

\textbf{SAS:} 
SAS received a significant number of comments regarding GUI and usability improvements, as well as learning and documentation, with 11 mentions for each category. This indicates that users frequently struggled with the complexity of the interface and the lack of intuitive guidance. P23 highlighted these issues by stating:

\begin{quote} \emph{"SAS could benefit from a more modern and intuitive graphical user interface (GUI). Currently, the interface feels outdated, and navigating through menus and options can be cumbersome for new users." } \end{quote}

This sentiment suggests that the interface may not meet current user expectations for software usability, potentially leading to a steep learning curve for new users. Performance issues were also a prominent concern among SAS users, with seven participants reporting instances of lag or software freezes during their tasks.

P17 described this problem as follows:

\begin{quote} \emph{``The software crashed part way through. A restart was necessary, which in turn removed all information that had already been processed and forced us to start from the beginning." } \end{quote}

The performance and stability issues not only disrupt workflow but also risk data loss, which can be particularly detrimental in data analysis tasks. 

\textbf{SPSS:}
For SPSS, the most commonly cited area for improvement was GUI and usability, with seven participants mentioning this in their responses. P39 expressed frustration with the interface complexity:

\begin{quote} \emph{"Too much information and buttons; had to go through a lot of tabs before getting the right one. Not UI/UX intrinsic." } \end{quote}

This feedback indicates that the abundance of features and options may overwhelm users, making it difficult to locate specific functions efficiently. P24 suggested that certain processes within SPSS could be streamlined to enhance usability:

\begin{quote} \emph{"I would like to see improved user interface simplicity in SPSS, as some processes like exporting data and conducting tests can feel unintuitive. Streamlining these steps with clearer guidance could enhance usability." } \end{quote}

These comments suggest that while SPSS is a powerful tool, its usability could be improved by simplifying the interface and making common tasks more accessible. Enhancements in these areas could reduce the learning curve for new users and improve overall efficiency.

\textbf{JMP:}

JMP received feedback indicating a need for improvements in GUI and usability, with 9 participants mentioning this aspect. Users expressed challenges with the clarity and intuitiveness of the interface. P41 commented:

\begin{quote} \emph{"The interface is not very clear and misleading in general. It is also impossible to copy just a single value from the output. The questions with how to modify the dataset are pretty confusing to implement using this software." } \end{quote}

This suggests that users found the interface confusing and had difficulty performing basic tasks, such as copying values and modifying datasets. These usability issues can hinder productivity and increase the learning curve for new users. Enhancing the clarity of the interface and simplifying common operations could significantly improve user satisfaction.

In addition to interface concerns, 3 participants mentioned the need for better error handling in JMP. P6 noted:

\begin{quote} \emph{"Better Error Messaging: Sometimes, error messages in JMP can be vague or unclear. Providing more specific, detailed error messages with suggestions for fixing issues would help users troubleshoot more effectively." } \end{quote}

This indicates that when users encountered errors, the lack of informative messages made troubleshooting challenging. Improving error messages to be more specific and actionable could facilitate a smoother user experience by enabling users to resolve issues more efficiently.

Despite these challenges, some participants reflected positively on JMP's improvements over other tools they had used. P14 stated:

\begin{quote} \emph{"The app shows significant improvements compared to the previous two apps [SAS, SPSS], demonstrating progress in functionality and user experience. However, there are still some issues that need to be addressed. While this app is better, it requires additional time and effort to navigate through errors effectively." } \end{quote}

This feedback acknowledges that JMP has made notable advancements in functionality and user experience compared to previously released tools such as SAS and SPSS. However, it also highlights that there are still areas, particularly in error navigation and overall usability, that require attention to meet user needs.

\textbf{Stata:}

Stata received the highest number of mentions for GUI and usability improvements among all the software evaluated, with 14 participants expressing this need. Users found the interface to be outdated and less intuitive compared to modern data analysis tools. P29 stated:

\begin{quote} \emph{"The Stata interface can feel dated compared to modern data analysis software. A more intuitive, streamlined interface with better navigation could make it more accessible, especially for new users." } \end{quote}

This suggests that the current design may impede user efficiency and discourage adoption by new users due to its lack of contemporary usability standards. Modernizing the interface could enhance accessibility and reduce the learning curve.

Additionally, five participants reported issues with the documentation and guidance provided by Stata. P44 shared her experience as follows:

\begin{quote} \emph{"I want to add documentation in the program, and guide before to start work, because I could not find how to work with it, then I googled and GPTed it." } \end{quote}

This indicates that the absence of built-in documentation or tutorials compelled users to seek external resources, such as online searches or AI assistance, to understand how to use the software. Providing comprehensive, obtainable on-demand learning within the program could streamline the learning process and improve user autonomy.

Despite these challenges, Stata also received positive feedback from participants. One user noted:

\begin{quote} \emph{"This software is more understandable compared to the previous ones. The errors displayed are also clear." } \end{quote}

This suggests that while there are areas for improvement, some users found Stata to be more comprehensible and appreciated the clarity of its error messages compared to other software like SAS and JMP. Enhancing the aspects that users already find favorable could further strengthen Stata's user experience.

\textbf{\textit{StatZ}:}

\textit{StatZ} stood out by receiving overwhelmingly positive feedback from participants. However, some users mentioned issues related to GUI and usability, with eight participants expressing a desire for improvements in this area. P19  commented:

\begin{quote} \emph{"I feel like there's too little information and application that you could do with this program compared to the other programs. Furthermore, I would like more images and UI improvement to help the user navigate around the program." } \end{quote}

This suggests that while the software is user-friendly, expanding its functionality and enhancing visual aids, such as tooltip, could further improve the user experience. Users may benefit from additional features and a more visually engaging interface to facilitate navigation and exploration of the software's capabilities.

Additionally, the detailed descriptions provided to assist users unfamiliar with the operations were perceived as cumbersome by more experienced users. P8 stated:

\begin{quote} \emph{"There are too many words on the homepage, maybe highlight what test or method can be used?" } \end{quote}

These types of comments indicate the need to balance the amount of explanatory text to cater to users with varying levels of expertise. Implementing a more streamlined layout or offering customizable views could allow users to access the information most relevant to their needs without feeling overwhelmed.

Despite these minor critiques, the majority of participants reflected positively and expressed enthusiasm for \textit{StatZ}. P12 stated:

\begin{quote} \emph{"It was definitely much easier than all previous applications! It was easy to follow, easy to get the information we need." } \end{quote}

P47 emphasized the user-friendliness of \textit{StatZ}:

\begin{quote} \emph{"I found this to be exceptionally user-friendly and overall very good. The interface is intuitive, making it easy to navigate and utilize effectively. I appreciate how seamlessly everything works together, which enhances the overall experience." } \end{quote}

This underscores the effectiveness of \textit{StatZ}'s design in facilitating an efficient and enjoyable user experience. The seamless integration of features and intuitive navigation appear to be key strengths appreciated by users.

Participants with less statistical background also found \textit{StatZ} particularly accommodating. P23 reflected:

\begin{quote} \emph{"I appreciated that everything was explained with \textit{StatZ}. I wish the layout was a little better, maybe a more pleasing GUI, but overall it was very easy to navigate. I think this tool was the best so far, as someone with very little statistical background." } \end{quote}

This suggests that \textit{StatZ} successfully lowers the barrier to entry for users with limited experience in statistics, offering clear explanations and ease of navigation. Enhancements to the visual layout could further improve the experience for novice users, making the software both accessible and aesthetically pleasing.

\section{Discussions and Findings}
User feedback, along with our quantitative results, indicate positive sentiments with \textit{StatZ}. Specifically, 16 users reported positive experiences with \textit{StatZ}, compared to only five users who rated Stata positively, which was the second-highest number of positive feedback. This stark contrast mainly underscores the influence of on-demand learning, reduced cognitive load, and other features offered by a conversational agent for statistical analysis. 

This section summarizes our interpretations of using a conversational agent for conducting statistical analysis. These findings are a design guideline for developers and individuals who are building data analysis tools.

\textbf{* Providing contextual guidance (on-demand learning) reduces the learning curve required for complex workflows.}

The integration of contextual explanations educates users and operates as a software wizard, guiding them to select the optimal function in complex workflows. In other words, these explanations explain the methodological rationale and map procedures to underlying statistical theory. Our empirical observations reveal that users of conventional statistical tools exhibit consistent seeking of external resources. Including documentation searches (86.3\% of participants) and AI-assisted consultations (92.2\% of participants) for basic operations guidance. This external dependency introduces multiple failure points, including AI model hallucination and temporal inefficiencies, caused by frequent switching between different interfaces. \textit{StatZ}'s architecture mitigates these risks through a guidance framework of 42 rigorously validated statistical methods. The closed-loop design eliminates dependencies on external verification mechanisms and reduced extraneous learning, as evidenced by decreased reliance and a faster completion duration.

\textbf{* The intuitive information filtering and minimalist interaction features of conversational agents mitigate cognitive load significantly.}

Our experiment revealed that participants using GUI-based tools frequently navigated complex menus to find the desired functionality. Additionally, we observed that participants often relied on AI assistance (LLMs) for navigation within these tools or for step-by-step guidance when using traditional statistical tools. This reliance on external resources and the need to experiment with various navigational features highlight the high cognitive load associated with traditional tools. On the other hand, the conversational agent includes intuitive task oriented information filtering, which enables users to concentrate on sequential actions with minimal distraction. In contrast, GUI tools often present numerous choices simultaneously, which can impose a higher cognitive load on users. This distinction is further supported by qualitative user feedback. A reduction in cognitive load correlates with improved accuracy and more efficient task completion. Specifically, users achieve correct results ($\Delta$acc = 219.1\%) with less time ($\Delta$t = -21.1\%) and effort compared to traditional GUI-based tools. With our design, users require significantly less effort to operate the system. \textit{StatZ} also demonstrates a significant decrease in keyboard interactions ($\Delta$\# keyboard = -61.23\%) and reduced mouse travel distance ($\Delta$Mouse Distance = -4.54\%) relative to the average of other tools. Minimizing motor demand translates into less physical effort for users, which enhances user satisfaction \cite{accot1997beyond}. This feature directly correlated to elevated Nelson usability score as shown in figure ~\ref{fig:userfeedback}.

\textbf{* The enforcement of proprietary formats negatively impacts usability by introducing unnecessary complexity into user interactions.}

% and unified file support are missing in traditional tools
Previously, proprietary statistical analysis tools dominated the market, enabling vendors to impose their proprietary file format convention. Nevertheless, the proliferation of different data analysis tools shifted the community towards adopting de-facto standard file formats such as CSV, TSV, XLXS, etc. While traditional tools support these files, their interface design often prioritizes proprietary formats, introducing usability friction that discourages the adoption of standardized alternatives. For instance, one of the simplest actions in data analysis is opening a file and loading the data. In tools such as \textit{StatZ}, users achieve seamless data loading via a single-step drag-and-drop interaction. On the other hand, GUI-based tools often impose multi-step workflow. such as SAS’s script-based file parsing or manual file-type selection from extensive menus. Empirical studies corroborate these challenges, eight users reported difficulty uploading and opening CSV data during their analysis. \textit{StatZ} elimintaes the necessity for users to save or export data in multiple proprietary formats (e.g., .jmp by JMP, .sas7bdat by SAS, .sav by SPSS, .dta by Stata); instead, unified file handling architecture enables direct drag-and-drop functionality, facilitating simplified exchange of data within and across collaborators. This capability mitigates file incompatibilities issues and reduces temporal overhead associated with dataset conversion, thereby boosting overall productivity. 

\subsection{Limitations}
Three principal limitations constrain the generalizability of our study. First, the sample size is limited to 51 healthy participants, which, while adequate for initial observations, may not completely represent the diversity of potential users. Our participants have no vision or physical disabilities, and we did not incorporate this into our study. This sampling bias limits ecological validity, as real-world users exhibit diverse cognitive and physical profiles. Second, the tasks assigned were limited to the most commonly employed statistical analyses, which may not encompass the full spectrum of statistical procedures used in various fields. The tools we compared our approach to offer a vast number of functionalities. However, due to the scientific nature of our work, we did not implement all of their features in \textit{StatZ}. Third, our study design did not capture long-term skill acquisition and learning effects that might emerge over the extended use of each software. These limitations underscore the need for future work to address inclusivity, methodological breadth, and temporal adaptability in our design.

\section{Conclusion and Future Work}
In this work, we introduced \textit{StatZ}, a conversational agent designed for statistical analysis. We have performed qualitative and quantitative comparisons between traditional statistical tools (SPSS, SAS, Stata, and JMP) with \textit{StatZ}. Our controlled experiment (N=51) measured three core metrics: task accuracy, task execution time, and user satisfaction (both quantitatively and qualitatively) across these platforms.  

Our findings indicate \textit{StatZ} users achieved significantly higher task accuracy (90\% vs 28.2\%, $p<0.01$) with 61\% fewer interactions, and completed tasks 26.6\% faster compared to those using traditional software. The conversational agent in \textit{StatZ} not only eliminates the need for external assistance but also reduces cognitive load by simplifying complex statistical concepts through intuitive, context-sensitive dialogue. While both traditional and conversational tools enable statistical analysis, the intuitive information filtering and on-demand learning capabilities of our \textit{StatZ} lead to an increase in user satisfaction, task completion time, and accuracy. Participants expressed a strong preference for the conversational interface, citing its alignment with usability principles such as clarity of feedback and effective error prevention. This preference suggests that conversational agents can bridge the gap between expert-only tools and those accessible to users with varying levels of statistical expertise. Additionally, removing the adherence to proprietary settings such as file format improves user experience by reducing the need for context switching. 

With ongoing technological advancements, understanding the human perspective and its influence on software design will be crucial in harnessing the full potential of conversational interfaces. While challenges remain, such as expanding the range of statistical procedures supported, there is significant potential in adopting a conversational style for statistical software design.

Future research should aim to include a larger and more diverse participant pool, incorporate a broader range of statistical tasks, and examine long-term user engagement and adoption. Embracing this approach may lead to more efficient workflows and democratization of data analysis across various disciplines.

\bibliographystyle{unsrtnat}
\bibliography{reference}

\appendix
\section{APPENDIX}
\subsection{Analysis Question}  
To establish the framework for our empirical investigation into the efficiency of statistical software, we consulted industry professionals specializing in data analysis and statistics to identify the most frequently utilized statistical operations in their daily workflows. Subsequently, we engaged three leading language models (ChatGPT\footnote{Developed by OpenAI, \url{https://openai.com/blog/chatgpt/}}, Meta.ai\footnote{Developed by Meta, \url{https://ai.facebook.com/}}, Claude.ai\footnote{Developed by Anthropic, \url{https://claude.ai}}) to compile a comprehensive list of these essential statistical operations. This dual approach ensured the relevance and comprehensiveness of the statistical tasks selected for our study. We then tasked a cohort participants to execute these operations across five different statistical analysis platforms—SPSS, SAS, Stata, and JMP, as well as through our interface. 

\begin{enumerate}
    \item \textbf{Data Importation}: Import the chosen dataset (Iris or NYC Taxi) into the software, ensuring proper data formatting and handling of any potential import issues.

    \item \textbf{Descriptive Statistics and Data Visualization}: Perform summarizing statistics such as mean, median, and standard deviation for specified variables. Participants were also asked to create basic plots, including histograms and scatter plots, to visualize data distributions and relationships between variables.

    \item \textbf{Inferential Statistics}: Conduct hypothesis testing relevant to the dataset. For the NYC Taxi dataset, for example, participants operated one-sample t-tests comparing the mean of the \textit{fare\_amount} variable to a sample mean, including reporting the p-value, t-statistic, and determining whether to reject the null hypothesis. They also performed independent t-tests to compare means between variables such as \textit{fare\_amount} and \textit{total\_amount}.

    \item \textbf{Normality Assessment}: Assessed whether certain variables were normally distributed using techniques such as Q-Q plots and the Shapiro-Wilk test, and report findings.

    \item \textbf{Correlation Analysis}: Calculated correlation coefficients between specified pairs of variables (e.g., \textit{trip\_distance}, \textit{tip\_amount}, and \textit{fare\_amount}) to examine the strength and direction of linear relationships.

    \item \textbf{Data Imputation}: Handle missing data by performing mean imputation on the original dataset and save the imputed dataset for submission.

    \item \textbf{Outlier Detection and Removal}: Detect outliers within the dataset using Isolation Forest algorithm \cite{liu2008isolation} and removed them, saving the cleaned dataset for submission.

    \item \textbf{Dimensionality Reduction}: Reduce the dimensionality of the dataset to a specified number of dimensions (e.g., reducing to four dimensions) using techniques such as tSNE\cite{van2008visualizing}, UMAP\cite{mcinnes2018umap} and PCA\cite{jolliffe2002principal}, and saved the reduced dataset.

    \item \textbf{Data Scaling}: Scale specified variables (e.g., \textit{tip\_amount}, \textit{fare\_amount}) using scaling methods such as min-max scaling to change a column and save the scaled dataset.

    \item \textbf{Data Exportation}: Save and export the processed data and results in specified formats for submission, ensuring that all outputs were correctly formatted and complete.
\end{enumerate}

These tasks were provided through online forms, with specific questions guiding the participants on what analyses to perform and what outputs to submit. For instance, participants analyzing the NYC Taxi dataset were asked to compute the mean of the \textit{fare\_amount} variable, perform t-tests, assess normality of \textit{trip\_distance}, compute correlation coefficients between variables, impute missing values, detect and remove outliers, reduce the dataset's dimensions, create scatter plots of specified variables, and scale certain variables. Participants uploaded results such as statistical outputs, plots, and processed datasets via the forms.

\subsection{Enhanced Prompt Design for Statistical Software Benchmarking}
To ensure the rigour and precision in testing the newly developed statistical analysis program, we adopt the structured Co-Star prompt engineering methodology. This technique, detailed in Sahoo's work \cite{Sahoo2023}, facilitates the creation of detailed and contextually rich prompts that guide the language model to generate specific, highly relevant responses.

\subsubsection{Co-Star Prompt Engineering Framework}
The Co-Star framework, as introduced by Sahoo \cite{Sahoo2023}, organizes prompt design into a structured format that enhances the effectiveness of language model interactions. This methodology involves specifying the context and the precise nature of the information required, which helps in tailoring the model's output to fit the user's exact needs. By leveraging this structured approach, prompts are crafted not only to solicit specific information but also to guide the language model through a logical progression of thought, mirroring human-like reasoning processes.

\subsubsection{Application to Statistical Software Benchmarking}
In the context of benchmarking a new statistical software against established programs, it is crucial to focus on core statistical operations that are universally recognized as essential for robust software evaluation. The following prompt utilizes the Co-Star format to ensure clarity and relevance, directing the language model to concentrate on fundamental statistical functions and exclude broader machine learning tasks.

\begin{quote}{\textbf{Prompt:}} {I am developing a new statistical analysis program and need to benchmark its capabilities against traditional statistical software such as SPSS, SAS, JMP. Could you provide a list of essential statistical operations typically used in such software for validation purposes? Please focus exclusively on statistical operations suitable for this comparison, excluding any machine learning tasks. The list should cover a broad range of functions including tests of means, variance analysis, regression models, and any other core statistical tools that are critical for a robust statistical software evaluation.} \end{quote}

This structured prompt directs the language model to generate a focused and detailed list of statistical operations, such as tests of means (t-tests, ANOVA), variance analysis (ANOVA, chi-square tests), and regression models (linear, logistic regression), which are critical in evaluating the effectiveness and accuracy of statistical software. The prompt deliberately excludes machine learning tasks to maintain the focus on traditional statistical methods, ensuring that the comparison remains relevant to the capabilities of the benchmarked software.

This approach aligns with the Co-Star methodology's emphasis on precision and context-specificity, which significantly enhances the utility of the language model's outputs for specialized tasks like software benchmarking.

\subsection{Analysis Question by Lagrange Model}

This appendix provides a comprehensive summary of the essential statistical operations provided by three different responses (ChatGPT, Meta, and ClaudeAI) for evaluating the capabilities of statistical analysis software.

The responses from ChatGPT, Meta, and ClaudeAI show a considerable overlap in their suggested statistical operations essential for software evaluation. All three models recommend foundational elements of statistical analysis, such as descriptive statistics with measures of central tendency and variability, and tests of means including one-sample, independent samples, and paired samples t-tests. Additionally, they each emphasize the importance of ANOVA (both one-way and two-way) and basic regression analysis (linear and logistic), which are critical for any statistical analysis toolkit. This commonality underscores these operations as universally essential for statistical analysis across various platforms.

However, each model also presents unique contributions that could serve specific analytical needs. ChatGPT distinguishes itself by including quality control methods and reliability testing, which are crucial for maintaining standards in production environments and psychological testing, respectively. Meta extends its utility by offering detailed data visualization tools and a suite of multivariate analysis techniques, making it exceptionally useful for complex data interpretation. ClaudeAI enhances its repertoire with advanced variance analysis techniques like MANOVA and ANCOVA and delves into distribution fitting and structural equation modeling, which are vital for more sophisticated statistical inquiries.

In summary, while there is a solid core of statistical operations endorsed by all three language models, their unique contributions demonstrate the diversity in their capabilities and potential applications. Organizations or individuals looking to benchmark or develop statistical software can leverage these insights to tailor their tools to specific needs or to ensure a comprehensive suite of functionalities that encompass both fundamental operations and advanced analytical techniques. This strategic selection of features can significantly enhance the robustness and applicability of statistical analysis software in various professional settings.

\subsubsection{ChatGPT's Statistical Operations}
\begin{enumerate}
    \item Descriptive Statistics: Measures of central tendency (Mean, Median, Mode); Measures of variability (Variance, Standard Deviation, Range, Interquartile Range).
    \item Tests of Means: One-Sample t-test, Independent Samples t-test, Paired Samples t-test.
    \item Analysis of Variance (ANOVA): One-way ANOVA, Two-way ANOVA, Repeated Measures ANOVA.
    \item Regression Analysis: Linear Regression, Multiple Regression, Logistic Regression.
    \item Non-parametric Tests: Mann-Whitney U Test, Kruskal-Wallis Test, Wilcoxon Signed-Rank Test.
    \item Correlation and Covariance: Pearson Correlation, Spearman's Rank Correlation, Covariance Analysis.
    \item Chi-Square Tests: Test for Independence, Goodness-of-Fit Test.
    \item Factor Analysis: Exploratory Factor Analysis, Confirmatory Factor Analysis.
    \item Quality Control: Control Charts, Pareto Analysis.
    \item Reliability Testing: Test-Retest Reliability, Cronbach's Alpha.
\end{enumerate}

\subsubsection{Meta's Statistical Operations}
\begin{enumerate}
    \item Descriptive Statistics: Mean, median, mode, Standard deviation, variance, range, interquartile range (IQR).
    \item Data Visualization: Histograms, box plots, scatter plots, bar charts.
    \item Inferential Statistics: Tests of Means (One-sample t-test, Two-sample t-test (independent samples), Paired t-test (dependent samples)); ANOVA (One-way ANOVA, Two-way ANOVA, Multi-way ANOVA); Variance Analysis (F-test for equality of variances, Levene's test for homogeneity of variances).
    \item Non-parametric Tests: Wilcoxon rank-sum test (Mann-Whitney U test), Wilcoxon signed-rank test, Kruskal-Wallis H test.
    \item Regression Models: Simple linear regression, Multiple linear regression, Polynomial regression, Logistic regression.
    \item Multivariate Analysis: Principal Component Analysis (PCA), Factor Analysis, Cluster Analysis (K-means clustering, Hierarchical clustering), Discriminant Analysis, Canonical Correlation Analysis.
    \item Hypothesis Testing: Z-test for proportions, Chi-squared test for independence, Fisher's exact test.
    \item Other Essential Statistical Tools: Data Transformation (Log transformation, Square root transformation, Standardization); Outlier Detection (Z-score method, Modified Z-score method, Density-based methods (e.g., DBSCAN)); Sampling Methods (Simple random sampling, Stratified sampling, Systematic sampling); Resampling Methods (Bootstrap sampling, Jackknife resampling).
\end{enumerate}

\subsubsection{Claude.ai's Statistical Operations}
\begin{enumerate}
    \item Descriptive Statistics: Measures of central tendency (mean, median, mode); Measures of dispersion (standard deviation, variance, range, interquartile range); Skewness and kurtosis; Frequency distributions and histograms.
    \item Tests of Means: One-sample t-test, Independent samples t-test, Paired samples t-test, One-way ANOVA, Repeated measures ANOVA, Factorial ANOVA.
    \item Nonparametric Tests: Mann-Whitney U test, Wilcoxon signed-rank test, Kruskal-Wallis test, Friedman test.
    \item Correlation Analysis: Pearson correlation coefficient, Spearman rank correlation, Kendall's tau, Partial correlation.
    \item Regression Analysis: Simple linear regression, Multiple linear regression, Polynomial regression, Stepwise regression, Logistic regression.
    \item Analysis of Variance (ANOVA) and Related Techniques: Two-way ANOVA, ANCOVA (Analysis of Covariance), MANOVA (Multivariate Analysis of Variance).
    \item Factor Analysis: Exploratory factor analysis, Principal component analysis.
    \item Reliability Analysis: Cronbach's alpha, Item-total correlation.
    \item Distribution Fitting and Other Statistical Tools: Normal distribution tests (e.g., Shapiro-Wilk, Kolmogorov-Smirnov), Chi-square goodness-of-fit test, Power analysis (Sample size calculation, Power calculation for various statistical tests), Contingency Table Analysis (Chi-square test of independence, Fisher's exact test, McNemar's test), Meta-analysis (Fixed-effects models, Random-effects models), Structural Equation Modeling (SEM) (Path analysis, Confirmatory factor analysis).
\end{enumerate}

\end{document}